\documentclass[twocolumn,prl,showpacs]{revtex4}
\usepackage{amsmath}%
\usepackage[latin1]{inputenc}%
\usepackage[dvips]{graphicx,color}
\usepackage{times}
\usepackage{here}

\newcommand{\Pd}[3]{\frac{\partial^{#1}#2}{\partial#3^{#1}}}
\newcommand{\PD}[2]{\frac{\partial#1}{\partial#2}}

\newcommand{\be}{\begin{equation}}
\newcommand{\ee}{\end{equation}}
\newcommand{\ie}{i.~e.\ }
\newcommand{\eg}{e.~g.\ }

\begin{document}
\title{New features of modulational instability of partially coherent light; \\importance of the incoherence spectrum}
\author{D. Anderson}
\author{L. Helczynski} \author{M. Lisak}\author{V. Semenov}\email{sss@appl.sci-nnov.ru}
\affiliation{Department of Electromagnetics, Chalmers University
of Technology, SE-412 96 Göteborg, Sweden} \affiliation{Institute
of Applied Physics RAS, 603950 Nizhny Novgorod, Russia}

\begin{abstract}
It is shown that the properties of the modulational instability of
partially coherent waves propagating in a nonlinear Kerr medium depend
crucially on the profile of the incoherent field spectrum.
Under certain conditions, the incoherence may even enhance, rather than suppress, the instability. In particular, it is found that the range of
modulationally unstable wave numbers does not necessarily
decrease monotonously  with increasing degree of incoherence and
that the modulational instability may still exist even when long
wavelength perturbations are stable.
\end{abstract}
\maketitle
\begin{figure}[h]
\includegraphics[scale=.43]{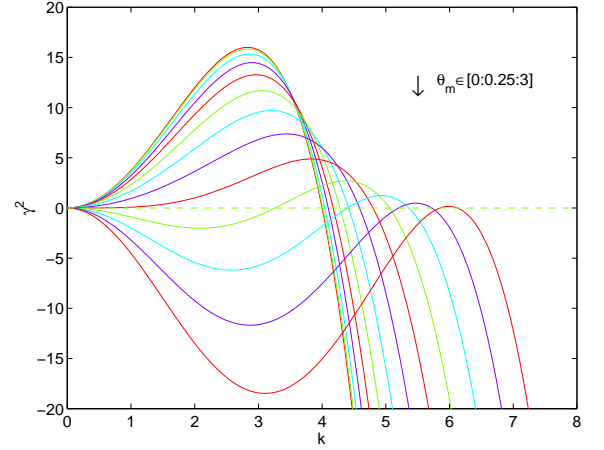}
\caption{The effect of increasing rectangular spectrum width
$\theta_m$ on the MI. The parameter $\theta_m$ runs from
$\theta_m=0$ (the top most curve) in increments of 0.25 to
$\theta_m=3$ (the bottom most curve). As can be seen, the
instability is never completely suppressed.
}\label{fig1:rectspec}\end{figure}
\begin{figure}[h]
\includegraphics[scale = .43]{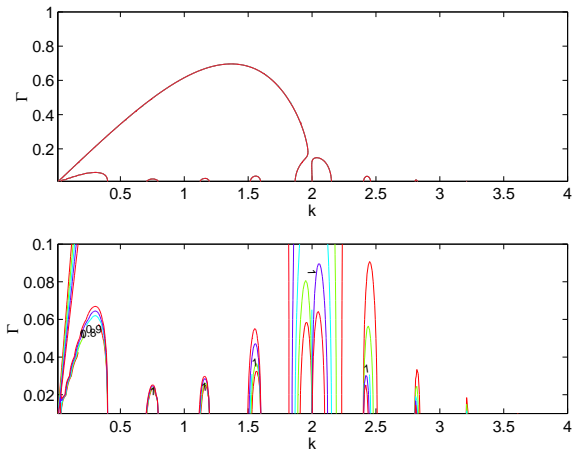}
\caption{Numerical solution of Eq.(\ref{realvalues}) for multi
carrier operation. The number of carriers is 10, $\alpha=0.2$, and
n$_0=0.8$. Top: Main growth curve together with the small
instability islands stemming from the separate carriers. Bottom: A
blow up of the instability curve. Notice the resonant peaks
occurring beyond the region of the main curve.  }\label{fig2:multicarrier}\end{figure}

\thispagestyle{plain} 
The modulational instability (MI) of coherent constant amplitude
waves in nonlinear Kerr media is one of the most fundamental
phenomena resulting from the interplay between nonlinear phase
modulation and linear dispersion/diffraction and has attracted
much interest over many years \cite{Karpman,Tanuti,Agrawal}. Recent advances in
the area of nonlinear optics, in particular new results
concerning the nonlinear propagation of partially incoherent light
and the advent of incoherent solitons \cite{Mitchell}, have
prompted a revisal of this issue during the past decade. The general understanding that emerged from these
studies is that the wave intensity threshold for the onset of the
MI is increased by the wave incoherence. With this picture in
mind, it is remarkable that in a recent investigation of the
transverse instability (TI) of solitons, \cite{Torres}, it was
found that the range of modulationally unstable wave numbers did
not monotonously decrease with increasing degree of incoherence.
In fact, it first increased until eventually it started to
decrease. \\Inspired by this result, we consider, in the present
work, the problem of the modulational instability of partially
coherent waves in more detail and show that the picture is more
complicated than previously thought. In order to simplify the
analysis and to bring out clearly the new features, the analysis
is carried out for the longitudinal modulational instability. We
find that the effect of the incoherence on the MI is sensitive to
the profile of the incoherent power spectrum. For the often used
assumption of a Lorentzian incoherence spectrum, the range of
unstable wave numbers does indeed decrease monotonously with
increasing degree of incoherence, whereas e.g. for a Gaussian
spectrum, the range first increases and then starts to decrease
monotonously. This result agrees well with the unexpected feature
observed in \cite{Torres}. Also, several other subtle effects are
shown to be possible. In particular, it is found that the
threshold for the MI to be completely quenched is not necessarily
associated with the long wavelength limit. Modulations may be
stable in this limit, but still be unstable for finite wave
numbers. This implies that the threshold for total quench can not,
in a general case, be determined by simplifying the analysis to
considering only the long wavelength limit as is done in \eg
\cite{Soljacic,Anastassiou}.\\
The starting point of our analysis is the normalized nonlinear Schrödinger equation describing the one dimensional propagation of a partially incoherent wave in a dispersive (or diffractive) nonlinear medium, viz \be i
\PD{\psi}{t}+\frac{1}{2}\Pd{2}{\psi}{x}+\langle|\psi|^2\rangle\psi=0\ee
where the bracket, $\langle\ldots\rangle$, denotes statistical
average \cite{Mietek}. This equation is valid under the assumption that the
medium response time is much larger than the characteristic time
of the stochastic intensity fluctuations.\\
The modulational instability of small perturbations of the
corresponding steady state solution has been analyzed using
different, but equivalent \cite{lukas}, mathematical formalisms.
An analysis based
on the formalism of the correlation function,
\cite{Wigner,Soljacic,Torres} or on the Wigner approach
\cite{Mietek}, results in the dispersion relation
\be\int^{+\infty}_{-\infty}\frac{\rho_0(p-k/2)-\rho_0(p+k/2)}{kp+i
\gamma}\,dp =1, \label{WignerDisp}\ee with $\rho_0$ being the
Wigner distribution function of the unperturbed cw wave. However, using the transformations:
$p+k/2=\theta$, $p-k/2=\theta'$, Eq.(\ref{WignerDisp}) can be
expressed as
\be\int^{+\infty}_{-\infty}\frac{\rho_0(\theta')\,d\theta'}{k(\theta'+k/2+i
\gamma/k)}-
\int^{+\infty}_{-\infty}\frac{\rho_0(\theta)\,d\theta}{h(\theta-k/2+i
\gamma/k)} =1\ee This is then easily rewritten as \be
k^2\int^{+\infty}_{-\infty}\frac{\rho_0(\theta)\,d\theta}{k^4/4+
(i  k \theta+\gamma)^2}=1\label{CDFDisp}\ee which is the same expression as
the Coherent Density approach \cite{Christo}, provided we identify
$\rho_0(\theta)=A^2\,G(\theta)$ with $A^2$ being the averaged
normalized field intensity of the stationary state and $G(\theta)$
being its normalized angular spectrum. Throughout this paper we will
use both expressions for the dispersion relation
interchangeably, since some parts of the analysis are most
conveniently handled by one approach, and some by the other.\\
An explicit analytical solution of the dispersion relation, Eq.(\ref{WignerDisp}) or Eq.(\ref{CDFDisp}), is possible only for
some particular incoherence spectra. Specifically, in the case
of the Lorentzian spectrum $G(\theta)=\theta_0/[\pi (\theta^2
+\theta_0^2)]$, where $\theta_0$ is the width of the spectrum, it has been shown \cite{Mietek}, that the
solution can be expressed as
\be \gamma(k,\theta_0)=\gamma_0(k)-k\theta_0 \label{gammak}\ee
where $\gamma_0(k)$ is the growth rate in the
coherent case ($G(\theta)=\delta(\theta)$) \ie
\be\gamma_0(k)=k\sqrt{A^2-k^2/4}, \label{gamma}\ee
with $A^2>k^2/4$. This analytical result shows explicitly that the effect of the
incoherence, provided it is large enough, is to suppress the MI for any
value of the perturbation wave number. However, as will be
demonstrated in this letter, this result depends crucially on the
form of the incoherence spectrum and is not a general feature of
the MI of partially incoherent light.\\
The restricted generality of the result expressed by
Eq.(\ref{gammak}) can be directly inferred by studying in more
detail the properties of the cut-off wave number, $k_c$, \ie the
value of $k$ at which the growth rate vanishes. According to
Eq.(\ref{gammak}), valid for the Lorentzian spectrum, $k_c$ is
shifted monotonously to the left (decreased) with increasing
degree of incoherence, $\theta_0$: $k_c^2=4(A^2-\theta_0^2)$. For
the case of a general coherence spectrum,  $k_c$ is determined by
the following equation:  \be p.v.\;\;
A^2\int^{+\infty}_{-\infty}\frac{G(\theta)d\theta}{
k_c^2/4-\theta^2}=1\label{PV}\ee where $p.v.$ denotes the
principal value. When the power spectrum is well localized, in the
sense that its rms-width is much smaller than the cut off
wavelength, the contributions from the
zeros of the denominator are negligible and the denominator can be
expanded to yield \be \frac{1}{A^2}\approx
\int^{+\infty}_{-\infty}\frac{G(\theta)}{
k_c^2/4}\left[1+\frac{\theta^2}{k_c^2/4}+\ldots\right]d\theta.
\label{approx}\ee Keeping only the first two terms of the
expansion one obtains an approximate solution for the effect of
partial coherence on the cut-off wave number as follows: \be
k_c^2/4\approx A^2+\langle\theta^2\rangle,\label{plus}\ee where
$\langle\theta^2\rangle\equiv\int\theta^2 G(\theta)d\theta$. Thus
$k_c$  is found to increase for increasing degree of incoherence,
which at first sight seems to be in contradiction to the behavior
of the previously found exact solution for the Lorentzian
spectrum, cf Eq.(\ref{gamma}). However, the Lorentz spectrum is
not well localized in the sense defined above, since the value
$\langle\theta^2\rangle$ does not exist. A first indication of
such an incoherence-induced increase of the range of
modulationally unstable wave numbers was observed by Torres et al.
\cite{Torres}, in a numerical study of the transverse instability
(TI) of soliton structures using a Gaussian spectral
distribution. It should be emphasized though, that the effect will
occur for transverse as well as for longitudinal modulational
instabilities. Although the Lorentz spectrum proper is not well
localized, it can easily be made so by considering the bounded
Lorentz spectrum \ie \be G(\theta,\theta_0,\theta_m)=
\frac{1}{\pi}\frac{\theta_0}{\theta^2+\theta_0^2}
\;C\;W(\theta_m-|\theta|) \ee where
$W(x)=0$ if $x<0$, $W(x)=1$
if $x>0$, $\theta_m$ is the  boundary of the spectrum and C is a
normalization constant. For this spectrum one can explicitly show
that, depending on the value of  $\theta_m$, the cut-off shift may
either increase or decrease with increasing $\theta_0$.\\
Another perturbative solution of the dispersion equation for the
MI is possible in the long wave limit when $k\ll\theta_0$, where
$\theta_0$ is the characteristic width of the spectrum
G($\theta$). In this case one can use a Taylor expansion  around
$p$ of the numerator in Eq.(\ref{WignerDisp}). Introducing
$\Gamma=\gamma/k$ we obtain \be
1=-A^2\int_{-\infty}^{+\infty}\frac{G'(p)}{p+i \Gamma}dp-
\frac{1}{24}A^2k^2\int_{-\infty}^{+\infty}\frac{G'''(p)}{p+i
\Gamma}dp.\label{elvan}\ee This simplified approximation (however
without the last term on the right hand side of Eq.(\ref{elvan}))
was used previously \cite{Soljacic} to analyze the threshold
condition for the suppression of the MI. Actually, assuming
$G(p)=G(-p)$ and $\Gamma=\Gamma^* \ll \theta_0$, the first
integral can be approximated as \be
\int_{-\infty}^{+\infty}\frac{G'(p)}{p+i \Gamma}dp= -J_1 +\pi
\Gamma D_1,\label{S1} \ee where \be
J_1=-\int_{-\infty}^{+\infty}\frac{G'}{p} dp,
\hspace{15mm}D_1=-\left.\frac{G'}{p}\right|_{p=0}. \label{S2} \ee
Within this approximation the dispersion relation becomes \be \pi
D_1 \Gamma = J_1 - A^{-2}. \label{S3} \ee The equality $J_1 =
A^{-2}$ can thus be taken as  determining  the threshold for MI
development. While this is true in the limit of vanishing $k$, it is
clear that by  taking into account also the next  term in the
expansion Eq.(\ref{elvan}), instead of Eq.(\ref{S3}) one obtains
the solution: \be \pi D_1 \Gamma = J_1 - A^{-2} - \frac{1}{24}k^2
J_3, \label{ADJ} \ee where \be
J_3=\int_{-\infty}^{+\infty}\frac{G'''}{p} dp. \ee This analysis
is valid provided $D_1\neq 0$, in other case a full solution of
Eq.(\ref{elvan}) is needed. Performing the calculations one finds
that for a Gaussian or a Lorentzian incoherence spectrum, the
values of $D_1$, $J_1$, and $J_3$ are all positive quantities.
Thus in such cases, Eq.(\ref{S3}) provides a sufficient condition
for the suppression of the instability, as found in the works of
Anastassiou et al. \cite{Anastassiou} and Solja\v{c}i\'{c} et al.
\cite{Soljacic}. Nevertheless, there may exist  spectra, for which
the factor $J_3$ is negative, implying there is positive
growth of the perturbation for finite k, despite the fact that
solution of the Eq.(\ref{S3}) (the long wave limit) gives
$\Gamma^2 \le 0$. Consequently, the definition for the suppression
of the modulational instability as the threshold value given by
the long wavelength limit approximation is not appropriate.\\
The simplest illustration of the ambiguity of the threshold
condition based on the long wavelength limit can be given by
analyzing a rectangular spectrum profile with $G(p) = 1/2\theta_m$
for  $p$ in the interval $-\theta_m<p<\theta_m$, whereas $G(p)
=0 $ outside this interval. For this simple spectrum, the integral
in Eq.(\ref{CDFDisp}) may be evaluated exactly and the following
dispersion relation is obtained \be
\gamma^2=k^2\left[\frac{k\theta_m}{\tanh(k\theta_m/A^2)}
-\frac{k^2}{4}-\theta_m^2\right]. \ee In the limit $k\theta_m\ll
A^2$, the dispersion relation can be approximated to read \be
\gamma^2\approx k^2\left[A^2-\theta_m^2-
k^2\left(\frac{1}{4}-\frac{\theta_m^2}{3A^2}\right)\right].\ee The
shift of the cut-off wave number $k_c$ is given approximately by\be
k_c^2\approx4(A^2+\theta_m^2/3),\ee provided $\theta_m\ll A$, which is in full agreement with
Eq.(\ref{plus}). However, the rectangular spectrum exhibits one
more unexpected and very important feature: It is evident that
even if $\gamma^2<0$ in the limit as $k\rightarrow 0$, $\gamma^2$
may become positive for finite $k$. In particular, when
$k=2\theta_m$, the growth rate of the perturbation is positive,
$\gamma^2>0$, independently of the spectrum width $\theta_m$, in
fact \be
\gamma^2=\frac{16\;\theta_m^4}{\exp(4\,\theta_m^2/A^2)-1}>0.\ee
Therefore, even a very high degree of incoherence does not
completely suppress the modulational instability when the spectrum
is rectangular, cf. Fig(\ref{fig1:rectspec}).
 Even  when $\theta_m\gg A$ and
the MI is strongly suppressed within the long wave range, there
still exists a "resonance" region of instability for wave numbers
$k$ around $2\theta_m$ given by : 
\be |k-2\theta_m| <
4\theta_m\exp\left(\frac{-\theta_m^2}{A^2}\right).\ee
It is interesting to note that a similar phenomenon was found in
\cite{Bang}, in an investigation of the effect of a nonlocal nonlinearity on the MI in a focusing Kerr medium. In particular, it was shown for a rectangular response function, that the instability growth rate first decreased for increasing width of the response function, but ultimately for large widths, new modulationally unstable bands appeared at finite wave numbers.\\
Since a rectangular profile is a very ideal and a bit artificial
form of the spectrum, we consider another example of a well
localized spectrum in the form of a modified Lorentzian, which
exhibits similar properties to those of a rectangular profile, \be
G(\theta)=\frac{\sqrt{2}}{\pi}\frac{\theta_0^3}{\theta^4+\theta_0^4}.\ee
Even in this case the integral of Eq.(\ref{CDFDisp}) can be
calculated in explicit form to yield
\begin{align}&\int^{+\infty}_{-\infty}\frac{G(\theta)d\theta}{(\gamma+i
k\theta)^2+(k^2/2)^2}= \\\nonumber
&\frac{k^4/4-\frac{k^2\theta_0^2}{2}+\left(\gamma+\frac{k\theta_0}
{\sqrt{2}}\right)^2+
2\left(\gamma+\frac{k\theta_0}{\sqrt{2}}\right)\frac{k\theta_0}{\sqrt{2}}}
{
\left[(\gamma+\frac{k\theta_0}{\sqrt{2}})^2+(\frac{k\theta_0}{\sqrt{2}}+\frac{k^2}{2})^2\right]
\left[(\gamma+\frac{k\theta_0}{\sqrt{2}})^2+(\frac{k\theta_0}{\sqrt{2}}-\frac{k^2}{2})^2\right]
}  \label{modLorDisp}
\end{align} Using this expression in the cut-off condition, $\gamma^2=0$,
we obtain the following result for the cut-off wave number: \be
k^2_c=2A^2\left\{ 1\pm\sqrt{1+4\left(\frac{\theta_0}{A}\right)^2
\left[1-\left(\frac{\theta_0}{A}\right)^2\right]}\right\}.\ee When
$1<(\theta_0/A)^2<(1+\sqrt{2})/2$, the equation has two positive
roots implying that the MI is not completely suppressed
provided this condition is fulfilled. On the other hand,
$\gamma^2>0$ only within a limited range of wave numbers, viz \be
\left|\frac{k^2}{2A^2}-1\right|<\sqrt{1+4\left(\frac{\theta_0}{A}\right)^2
\left[1-\left(\frac{\theta_0}{A}\right)^2\right]}.\ee\\
In order to further illustrate the subtlety of the interplay between the partial incoherence and the instability drive, we consider a multi-carrier case where the
field consists of many mutually incoherent, but individually partially coherent waves with a spectrum given by $G(\theta)=\sum_n G_n(\theta-\theta_n)$.
The dispersion relation then becomes \be
A^2k^2\sum_n\int_{-\infty}^{+\infty}\frac{G_n(\theta-\theta_n)
\,d\theta}{(\gamma +i k\theta)^2+(k^2/2)^2}=1.
\label{multiDisp}\ee For simplicity we consider the particular
case of equally separated carriers, \ie $\theta_n=\alpha n$ where
$n$ is an integer, with each carrier having a Lorentzian phase
spectrum of the same width \be
G_n(\theta-\theta_n)=\frac{g(n)\,\theta_0}{\pi[(\theta-\alpha
n)^2+ \theta_0^2]}.\ee The dispersion relation,
Eq.(\ref{multiDisp}), can then be written as  \be A^2k^2\sum_{n}
\frac{g(n)}{(\Gamma+i k \alpha n)^2+k^4/4}=1, \ee where
$\Gamma=\gamma+k\theta_0$. For a symmetric spectrum, i.e. when
$g(k)=g(-k)$, this dispersion relation can be rewritten in terms
of real functions: \be A^2\sum_{n}g(n)\frac{X+
(a+n\alpha)(a-n\alpha)}{[X+(a+n\alpha)^2]
[X+(a-n\alpha)^2]}=1\label{realvalues}\ee where $X=\Gamma^2/k^2$,
$a=k/2$ and the coefficients $g(n)$ are normalized to unity, $\sum_n
g(n)=1$. When $X$ has a small positive value, the sum in
Eq.(\ref{realvalues}) has multiple resonant values at $a_n=\alpha
n$, $n\neq 0$. Consequently, one should expect the existence of
small positive roots $X$ of Eq.(\ref{realvalues}) in the vicinity of
those resonances, independently of the structure of the spectrum
envelope $g(n)$. That this indeed is the case is illustrated in
Fig.(\ref{fig2:multicarrier}), which shows the result of a numerical evaluation of the sum for particular parameter values.
On the other hand, outside of these resonant bands, the summation
can be transformed into an integration over n provided the
spectrum $g(n)$ is dense enough \ie $\alpha\ll A$,
$|g(n+1)-g(n)|\ll g(n)$; \be A^2\int^{+\infty}_{-\infty} g(n)\frac{X+
(a+n\alpha)(a-n\alpha)}{[X+(a+n\alpha)^2][X+(a-n\alpha)^2]}dn=1.\label{Eq31}\ee
This equation coincides with Eq.(\ref{CDFDisp}) if $g(n)\,dn$ is
replaced by $G(\theta)\,d\theta$ and $n \alpha$ is changed to
$\theta$. It corresponds to the MI of a continuous spectrum which
coincides with the envelope of the actual spectrum. For example,
when \be g(n)=\frac{1}{\pi}\frac{n_0}{n^2+n_0^2},\ee where $n_0\gg
1$, Eq.(\ref{Eq31}) is reduced to the following well known
expression:
\be (\Gamma+k\alpha n_0)^2=k^2(A^2-k^2/4).\ee Thus,
within this approximation, the MI is suppressed for all wave
numbers provided $\alpha n_0>A$. However, as shown above, this is
correct only outside the resonance bands, i.e. the localized
regions around each $2\alpha n$ ($n\neq 0$), where $\Gamma^2>0$, independently of the width of the envelope ($n_0$). \\
To conclude, we have investigated the role of the incoherent
spectrum profile on the properties of the modulational
instability. The gain curve of the instability may smoothly shrink
in amplitude and cut-off wave number with increasing degree of
incoherence, as is the case for a Lorentzian profile. However, it
may also  initially expand into the wave numbers which are stable
in the coherent regime as is the case for a Gaussian profile of
the spectrum. When the spectrum is rectangular, we have shown that
under the special resonance condition $k=2\theta_m$, the MI cannot
be suppressed completely no matter how strong the degree of
incoherence. Using the modified Lorentzian profile we demonstrate
that the long wavelength threshold definition at which the MI is
suppressed due to partial incoherence is in fact profile
dependent. For this case, islands of positive growth rate emerge at
higher wave numbers. Lastly, our analysis of the modulational instability in the case of multi-carrier
operation demonstrates new additional
structure in the gain curve, where besides the main lobe, there also
exist smaller peaks  surrounding the discrete carrier phases.
\end{document}